\documentclass[prl,twocolumn,letterpaper,showpacs,superscriptaddress]{revtex4}
\usepackage{graphicx,bm,amssymb,amsmath}
\usepackage{color}

\def\be{\begin{equation}}
\def\fin{\end{equation}}

\def\T{{\sf T\kern-.45em T}}
\def\C{\kern.1em{\raise.47ex\hbox{$\scriptscriptstyle |$}}
       \kern-.40em{\sf C}}

\begin{document}
\title{Spontaneous contractility--mediated cortical flow generates cell migration in  3-dimensional environments}
\date{\today}
\author{ R.J. Hawkins}
\affiliation{UMR 7600, Universit\'e Pierre et Marie Curie/CNRS, 4 Place Jussieu, 75255
Paris Cedex 05 France}
\author{R. Poincloux}
\affiliation{UMR 144, Institut Curie/CNRS, 
26 rue d'Ulm 75248 Paris Cedex 05 France}
\author{ O. B\'enichou}
\affiliation{UMR 7600, Universit\'e Pierre et Marie Curie/CNRS, 4 Place Jussieu, 75255
Paris Cedex 05 France}
\author{M. Piel}
\affiliation{UMR 144, Institut Curie/CNRS, 
26 rue d'Ulm 75248 Paris Cedex 05 France}
\author{P.Chavrier}
\affiliation{UMR 144, Institut Curie/CNRS, 
26 rue d'Ulm 75248 Paris Cedex 05 France}
\author{ R.Voituriez}
\affiliation{UMR 7600, Universit\'e Pierre et Marie Curie/CNRS, 4 Place Jussieu, 75255
Paris Cedex 05 France}
%\affiliation{UMR 7600, Universit\'e Pierre et Marie Curie/CNRS, 4 Place Jussieu, 75255
%Paris Cedex 05 France}

%\author{Rhoda J. Hawkins}
%\affiliation{UMR 7600, Universit\'e Pierre et Marie Curie/CNRS, 4 Place Jussieu, 75255
%Paris Cedex 05 France}
%\author{M. Piel}
%\affiliation{UMR 144, Institut Curie/CNRS, 26 rue d'Ulm 75248 Paris Cedex 05 France}
%\author{Rapha\"el Voituriez}
%\affiliation{UMR 7600, Universit\'e Pierre et Marie Curie/CNRS, 4 Place Jussieu, 75255
%Paris Cedex 05 France}

%\pacs{87.10.-e, 87.17.Jj, 83.80.Lz}%{Biological physics, theory}

\begin{abstract}
 We present a  generic model of cell motility generated by acto-myosin contraction of the cell cortex.  We identify analytically  dynamical instabilities of the cortex and show that they trigger spontaneous cortical flows which in turn can induce cell migration in 3-dimensional (3D) environments as well as bleb formation. This contractility--based mechanism, widely independent of actin treadmilling,  appears as an alternative to the classical picture of lamellipodial motility on flat substrates. Theoretical predictions are compared to experimental data of tumor cells migrating in 3D matrigel and suggest that this mechanism could be a general mode of cell migration  in 3D environments.
\end{abstract}

\maketitle

Beyond its clear interest in the context of cell and molecular biology \cite{alberts},  the study of cell motility  and more generally the understanding of
simple mechanisms of self-propelled motion is an
important challenge for physics and biomimetic technology. Within this context, identifying mechanisms of migration of micro-organisms and in particular living cells has motivated numerous works in the biology and physics communities \cite{Noireaux:2000ia,Bernheim-Groswasser:2002xa}. 
Sustained motion at low Reynolds number
%, the relevant regime at the cell scale, 
necessitates a constant energy input, and
therefore an active system, that is a system driven out of equilibrium
by an internal or an external energy source.   The cell cytoskeleton has been long identified as 
 an example of such an active system. It is a network of
semi-flexible filaments made up of protein subunits, interacting with
other proteins 
%which can, among other things, crosslink or cap the
%filaments\cite{alberts}. M
such as 
motor proteins
%, myosins, kinesins or dyneins
which use chemical energy 
of ATP hydrolysis 
to  exert active stresses that
deform the network \cite{alberts}.

Two  principal out-of-equilibrium processes have been identified as responsible for cell motility: polymerization (treadmilling) of cytoskeleton filaments and contractility of the cytoskeleton, which results from the interaction between actin filaments and motor proteins. Both polymerization induced motion \cite{Noireaux:2000ia,Bernheim-Groswasser:2002xa} and contractility induced spontaneous
flows \cite{Nedelec1997}  have now been observed in-vitro, and studied
theoretically \cite{Kruse2004,Voituriez2005}
and numerically \cite{Marenduzzo:2007}. In all models the key ingredients of motion are an energy input
to compensate dissipation and sufficient adhesion or
friction with a substrate to acquire momentum. The usual
picture of cell locomotion, following classical motility experiments realized on 2-dimensional (2D) flat substrates,   is then as follows: the cell lamellipodium builds strong adhesion
points with the substrate and pushes  its membrane forward by
polymerizing actin. At the back, the cell body contracts and breaks
the adhesion points \cite{alberts,Yam:2007ao}. % In particular, the overall cell velocity is then
%limited by the actin polymerization rate (which however varies substantially between cell types), in agreement with 
%available experimental data  \cite{Theriot1991,Clainche2008,Pollard2003}.

\begin{figure}[hbt]
\begin{centering}
\includegraphics[width=0.45\textwidth]{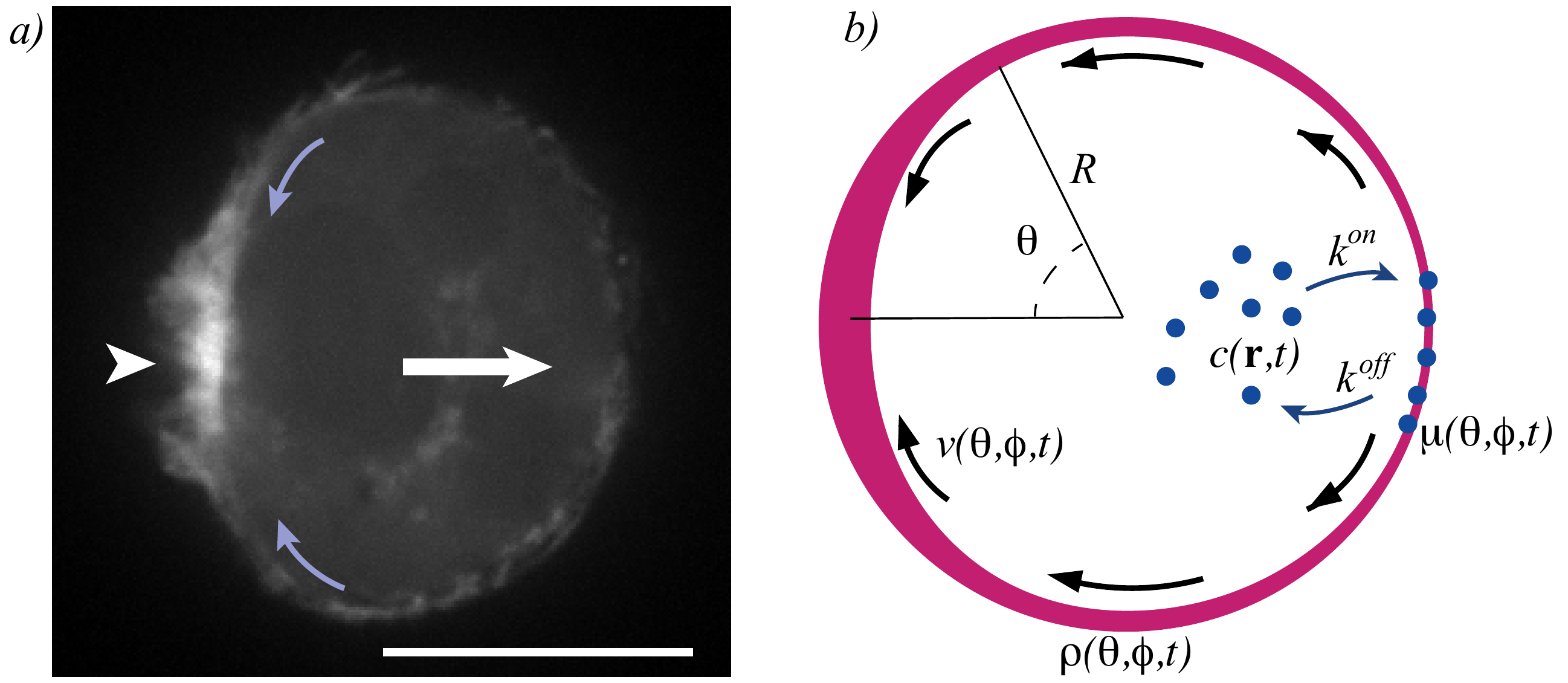}
\caption{\label{fig_geom} Cell migration in 3D environments. a) Confocal section of an MDA-MB-231 cell  migrating in 3D Matrigel. The cell expresses mCh-Lifeact, a fluorescent probe that labels F-actin \cite{Poincloux2010}. White arrow indicates the direction of movement. Arrowhead points to the accumulation of F-actin at the rear of the cell. Blue arrows indicate the direction of observed cortical flow analyzed in Fig. \ref{fig_donnevitesse}. Scale bar, 10 $\mu m$. b) Cartoon of the model in the first unstable mode $Y_{1,0}$, with the actin cortex in purple.  Arrows indicate the velocity of the gel.}
\end{centering}
\end{figure}

However, the geometry of the cell environment in vivo is significantly different from a flat substrate. The effect of geometry -- and in particular confinement --  has been shown to play a crucial role in cell migration, enabling the use of mechanisms widely different from lamellipodial motility \cite{Lammermann:2008tf,Hawkins2009}. In this context, studying cell motility in 3D environments is a promising and widely unexplored field.   %As suggested by \cite{Charras2008} and others the phenomenon of cell blebbing, generated by the contraction of the actin cortex,  may be used by the cell to aid in its migration. 
In fact, recent observations \cite{Poincloux2010} reveal that MDA-MB-231 breast tumor cells migrate in 3D matrigel with a spherical shape (see Fig. \ref{fig_geom}) according to a contraction-based mechanism in absence of actin protrusion or lamellipodium formation at the leading edge. 
Here, inspired by these experiments,  we develop a generic model for motility based on the active contraction of the cell's actin cortex caused by the action of myosin. We analytically identify dynamical instabilities of the cortex and demonstrate that spontaneous cortical flows appear. We show that such cortical flows, observed in many contexts of cell polarization and development  \cite{Bray:1988uq,Mayer:2010fk}, can also induce cell migration in 3D environments. Theoretical predictions are compared to experimental data of \cite{Poincloux2010}, and  more generally suggest that this contractility--based mechanism could be a generic mode of migration in 3D environments which strongly differs from the classical picture of lamellipodial motility.

The model is as follows (see Fig. \ref{fig_geom}). For the sake of simplicity we consider a  cell of spherical shape and conserved volume with radius $R$, and use spherical coordinates $(r,\theta,\phi)$ (qualitatively similar results are expected for other geometries).
 The cell cortex is a thin shell of acto-myosin gel which  polymerizes at the membrane (which defines the outer boundary at $r=R$) with speed $v_p$, and depolymerizes with a constant rate $k_d$.  Since the actin cortex can vary in thickness, it can be  modeled effectively as a 2D shell of  a compressible gel with density $\rho(\theta, \phi,t)=\rho_0(1+\delta\rho(\theta,\phi,t))$. The average density $\rho_0$ is set by the polymerization speed $v_p$ according to  $\rho_0=av_p/k_d$ where $a$ is the actin monomer concentration which for simplicity we assume is constant. 
 Mass conservation is then given by Eq. (\ref{eq3'}) below. In a first approximation we write the pressure in the cortex $P=\alpha\delta\rho-\beta\nabla^2\delta\rho$, where $\alpha^{-1}$ is the compressibility of the  gel and $\beta$ the correlation length of density fluctuations, and denote ${\bold v}=(v_\theta,v_\phi)$ the components of the gel velocity. Here, we consider only irrotational flows in the cortex and therefore write $\bold v=\nabla \psi$. 
Hereafter the operator $\nabla$ implicitly acts on the angular variables $\theta,\phi$. We neglect the effects of actin polarization and assume that the gel is in an  isotropic phase. The activity of myosin motors then results in a diagonal active stress $\sigma^a _{\alpha\beta}=-\zeta\mu\delta_{\alpha\beta} $ in the gel which, following \cite{Salbreux2007},  we assume is proportional to the local concentration of myosin in the cortex  $\mu(\theta,\phi, t)$,  where $\zeta$ is a phenomenological coupling. In the low Reynolds number regime, which is the relevant regime at the cell scale, the force balance given by Eq.   (\ref{eq4'}) is satisfied,   where we denote by $\xi$  the friction with the external medium. Here we assume for the sake of simplicity that viscous effects in the gel are much smaller than friction and can be neglected as assumed in \cite{Callan-Jones:2008sf}. The case of viscosity dominated dynamics, observed in \cite{Mayer:2010fk}, can be studied along the same line and yields qualitatively similar results.
In the cell bulk actin filaments are much more diluted than in the cortex and assumed to be in an isotropic phase, so that  myosin, whose bulk concentration  is denoted by $c({\bold r},t)$,  diffuses  with diffusion constant $D_c$ (see Eq. (\ref{eq0'}) where $\Delta$ denotes the 3D Laplacian). The myosin motors can  attach to the cortex at rate $k^{\text{on}}$, where they are transported with the flow of the actin cortex itself and diffuse  with diffusion constant $D_{\mu}$, and can fall off from the cortex with rate $k^{\text{off}}$. Eq. (\ref{eq1'}) describes the conservation of myosin at the cortex/cytoplasm interface, and  Eq. (\ref{eq2'}) in the cortex. Finally, the non linear dynamical equations for the system are written:
\begin{align}
\label{eq3'}
&\partial_t \delta\rho+\nabla\cdot \left ((1+\delta\rho)\nabla \psi\right )=-k_d\delta\rho\\
\label{eq4'}
&\nabla\left (-\zeta\mu-\alpha\delta\rho+\beta\nabla^2\delta\rho\right )=\xi \nabla \psi\\
\label{eq0'}
& \partial_t c\;=D_c\Delta c  \\
\label{eq1'}
&-D_c \partial_r c\lvert_{r=R}\;=k^{\text{on}}c(R)-k^{\text{off}}\mu\\
\label{eq2'}
&\partial_t \mu+\nabla\cdot (\mu \nabla \psi)=k^{\text{on}}c(R)-k^{\text{off}}\mu+D_{\mu}\nabla^2\mu.
\end{align}

\begin{figure}[hbt]
\begin{centering}
\includegraphics[width=0.42\textwidth]{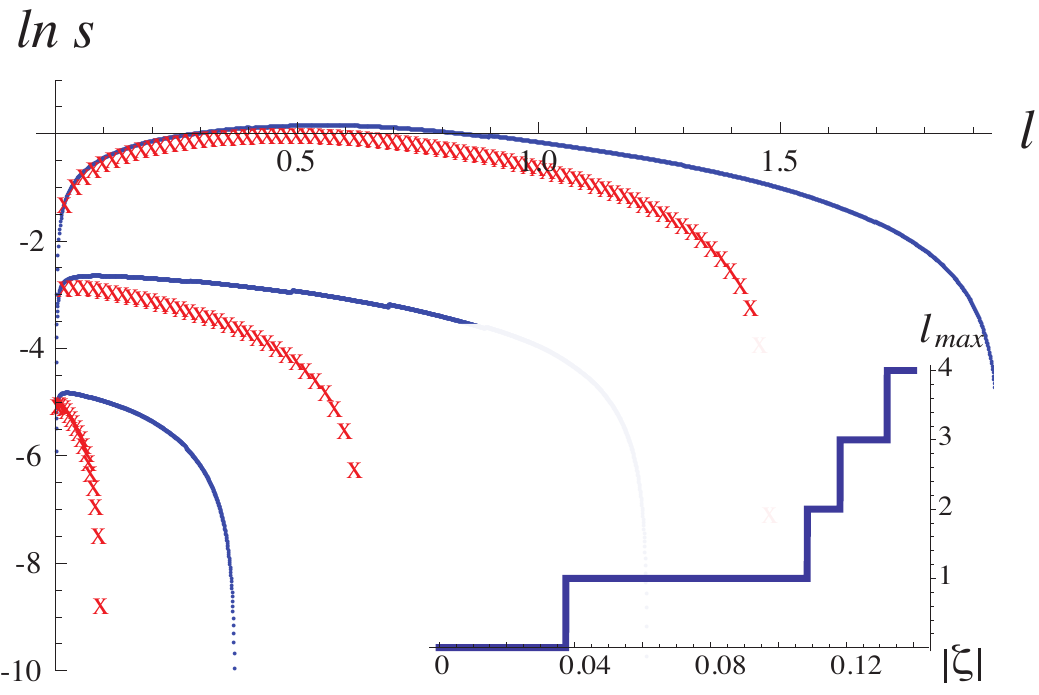}
\caption{\label{fig_s(l)activity} Dispersion relation $s(l)$ in log scale: blue lines 1D (where $k=\sqrt{l(l+1)}/R$) and red crosses 3D. Pairs of curves in ascending order are for activities $\zeta=-0.01$, $-0.05$, $-0.1\,{\text{kg}}\,\mu{\text{m}}\,{\text{s}}^{-2}$. Inset :  most unstable integer mode number $l_{\text{max}}$ (integer $l$ when $s(l)$ is maximal), against the strength of activity $|\zeta|$. Other parameter values (order of magnitude estimates of biologically relevant values taken from \cite{Mogilner2002,Pollard2000,Hawkins2009a,Callan-Jones:2008sf,Wagner1999,Niederman1975,Kolega1993}): $k^{\text{on}}=1\,\mu{\text{m}}\,{\text{s}}^{-1}$, $k^{\text{off}}=0.1\,{\text{s}}^{-1}$, $D_c=10\,\mu{\text{m}}^{2}\,{\text{s}}^{-1}$, $D_{\mu}=1\,\mu{\text{m}}^{2}\,{\text{s}}^{-1}$, $k_d=0.1\,{\text{s}}^{-1}$, $\alpha=1000\,{\text{kg}}\,\mu{\text{m}}^{-1}\,{\text{s}}^{-2}$, $\beta=1000\,{\text{kg}}\,\mu{\text{m}}\,{\text{s}}^{-2}$, $\xi=0.1\,{\text{kg}}\,\mu{\text{m}}^{-3}\,{\text{s}}^{-1}$, $R=10\,\mu{\text{m}}$ and $\mu_0=10^{4}\,\mu{\text{m}}^{-2}$.}
\end{centering}
\end{figure}

The homogeneous static solution of Eq. (\ref{eq3'}-\ref{eq2'}) is trivially given by $\psi=\delta\rho=0$, $\mu=\mu_0$ and $c=k^{\text{off}}\mu_0/k^{\text{on}}$. In this case there is no actin flow in the cortex and the cell is at rest. To determine the linear stability of this homogeneous state we consider a perturbation of the $(l,m)$ spherical harmonic  for each variable.  Namely  we let $c(r,\phi)=c_0+c_{l,m}(r)Y_{l,m}(\theta,\phi)e^{st}$ where $c_0$ is the homogeneous static solution given by $c_0=k^{\text{off}}\mu_0/k^{\text{on}}$,  and define along the same lines the other $(l,m)$ components of the perturbation as  $\psi_{l,m}, \delta\rho_{l,m},\mu_{l,m}$. Solving Eq.~(\ref{eq0'}) then gives $c_{l,m}(r)=\tilde c_{l,m}I_{l+1/2}(\sqrt{\frac{s}{D_c}}r)/\sqrt{r}$ in terms of the Bessel function $I_n$.   We then obtain the following linearized homogeneous system: 

\begin{align}
\label{eq3bis}
& s\delta\rho_{l,m}+k_d\delta\rho_{l,m}-\frac{l(l+1)}{R^2}\psi_{l,m}=0\\
\label{eq4bis}
&-\zeta\mu_{l,m}-\alpha\delta\rho_{l,m}-\beta \frac{l(l+1)}{R^2}\delta\rho_{l,m}-\xi \psi_{l,m}=0\\
\label{eq1bis}
& D_c\left((l+1-k^{\text{on}}R/D_c)I_{l+1/2}(u)-uI_{l-1/2}(u)\right)\tilde c_{l,m}\\\nonumber
&+k^{\text{off}}R^{3/2} \mu_{l,m}=0\\
\label{eq2bis}
&(s+k^{\text{off}}+\frac{l(l+1)}{R^2}D_\mu) \mu_{l,m}-\mu_0 \frac{l(l+1)}{R^2}\psi_{l,m}\\\nonumber
&-k^{\text{on}}\tilde c_{l,m}I_{l+1/2}(u)/\sqrt{R}=0
\end{align}
where $u\equiv\sqrt{\frac{s}{D_c}}R$.  Demanding a non zero solution yields an explicit equation for $s$ which defines the dispersion relation of the problem.
This equation, which is non polynomial,  can be studied numerically and fully characterizes the linear dynamics of the problem.
A simplified (and analytically tractable) dispersion relation can be obtained in the limit where curvature effects can be neglected, which essentially amounts to considering the equivalent 1D problem. This assumption is valid at  time scales $\tau\gg R^2/D_c$, when the bulk concentration of myosin can be taken as constant. It can be shown that in this regime the dispersion relation is cubic in $s$ and reads 
\begin{align}\label{disp1d}
\Big ((s+D_ck^2)&(s+k^{\text{off}}+D_{\mu}k^2)+k^{\text{on}}(s+D_{\mu}k^2)\Big )\nonumber\\ &\times\Big (\xi(s+k_d)+ k^2(\alpha+\beta k^2)\Big )
+\nonumber\\ &\zeta\mu_0k^2(s+k_d)(s+k^{\text{on}}+D_ck^2)=0,
\end{align}
where $k^2\equiv l(l+1)/R^2$.
A numerical analysis shows that for the biologically relevant range of parameters (see figure \ref{fig_s(l)activity}), this 1D approximation is quantitatively very close to the exact dispersion relation, which means that in practice curvature effects can be  neglected.  This 1D approximation  (\ref{disp1d}) is useful   to  determine the threshold activity beyond which the instability appears. Indeed, a straightforward analysis  shows that beyond the critical value  defined by 
\begin{equation}
\zeta_c =-\frac{(D_ck^{\text{off}}+D_{\mu}k^{\text{on}})\xi}{\mu_0k^{\text{on}}}
\end{equation}
real positive solutions for $s$ exist, which means that the homogeneous state
is unstable and that cortical flows appear. This threshold is controlled by
both the diffusion properties of myosin, which clearly stabilize the
concentration profile, and the friction which governs the amplitude of the
cortical flow. Importantly, this threshold is independent of the physical
properties of the gel, $\alpha$ and $\beta$, which however enter in
stabilizing terms determining the shape of the unstable pattern. Moreover, the
analysis of Eq. (\ref{disp1d}) reveals that $s$ is maximized for a finite
mode $l_{\rm max}$ which will be dominant to linear order and will
characterize the developing flow pattern at least at short times. Figure
(\ref{fig_s(l)activity}) shows that the most unstable mode $l_{\rm max}$ grows
with the activity. Note  that a full non linear treatment, which goes beyond
the scope of this paper,  would be necessary to discuss the long time behavior
of the problem. Interestingly, the specific limit of fast exchange and fast
depolymerization  ($k^\text{on},k^\text{off},k_d$ large and $D_c=0$ for
simplicity) can be discussed at non linear order. In this case the dynamics takes the simplified form 
\begin{equation}
\partial_t \mu= D_\mu\nabla^2\mu+\frac{\zeta}{\xi}\nabla\cdot(\mu\nabla\mu),
\end{equation} 
which can be viewed as a modified Keller-Segel equation with drift $-\zeta/\xi\nabla\mu$ \cite{Keller1971}. In this limit it can be shown that the solution blows-up as soon as $\mu>-\xi  D_\mu/\zeta$. We expect that in the physical regime this singularity is damped leading to smoother asymptotic profiles.
\begin{figure}
\begin{centering}
\includegraphics[width=0.45\textwidth]{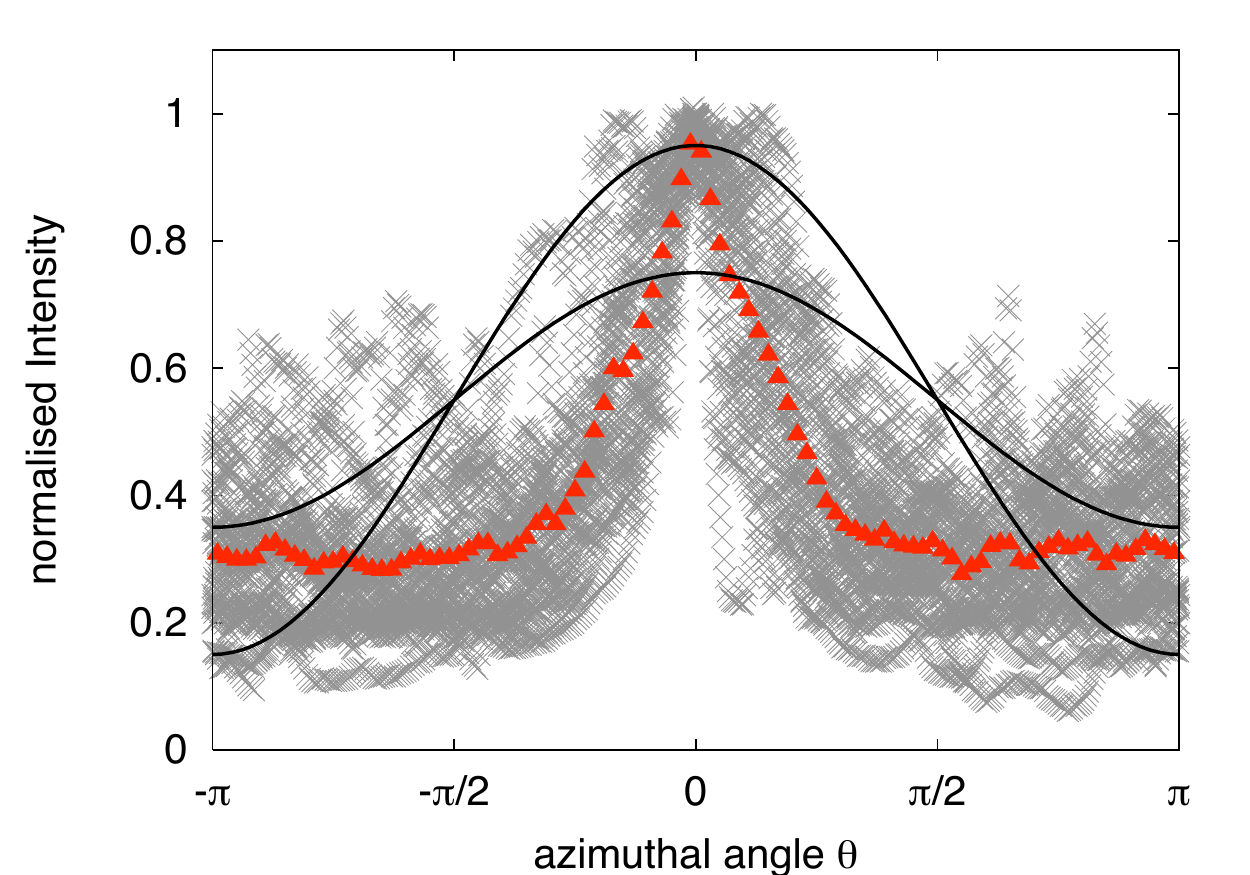}
\caption{\label{fig_donneintensite}Experimental data from \cite{Poincloux2010}
  for the intensity of mCherry-Lifeact labeled actin around the cortex of MDA-MB-231 tumor  cells
 seeded in 3D Matrigel. Theoretical line in black shows
  the predicted unstable mode $Y_{1,0}(\theta)$. Red triangles show the
  average of 19 individual cells shown with grey crosses. The position $\theta=0$ is set at the maximum intensity along the perimeter and corresponds to the back of the moving cells.}
\end{centering}
\end{figure}
\begin{figure}%[hbt]
\begin{centering}
\includegraphics[width=0.45\textwidth]{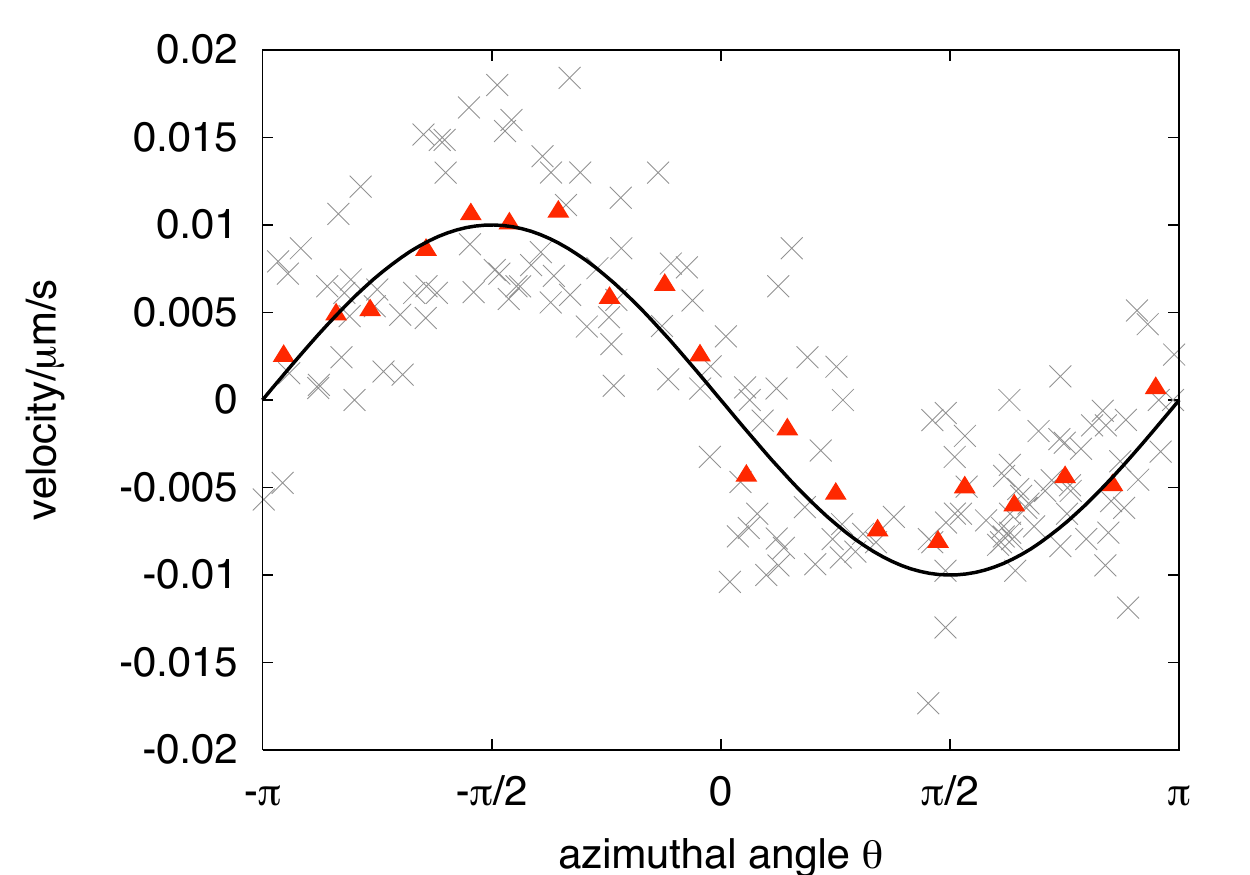}
\caption{\label{fig_donnevitesse}Experimental data from \cite{Poincloux2010}
  for the velocity of actin around the cell cortex calculated from kymographs
  (position over time) of the data in figure \ref{fig_donneintensite}. Red
  triangles show the average of 9 individual cells shown with grey crosses. The same reference $\theta=0$ as in figure \ref{fig_donneintensite} applies, with the black line showing the theoretical velocity of the predicted unstable mode $Y_{1,0}(\theta)$.}
\end{centering}
\end{figure}

These results have several implications in the context of cell biology.
Firstly, we show that such spontaneous cortical flows, which are observed in many contexts of cell polarization and development \cite{Bray:1988uq,Mayer:2010fk},  can induce motion when the cell is embedded in a 3D medium. Quantitatively, the above analysis shows that for $\zeta>\zeta_c$  the mode $l=1$, for which $\mu,\delta\rho, \psi$ are proportional to $Y_{1,0}(\theta)$, grows and creates a converging flow towards the pole $\theta=0$ of the cell where the cortex  thickens, while the pole $\theta=\pi$ depletes, as shown in figures (\ref{fig_geom},\ref{fig_donneintensite},\ref{fig_donnevitesse}).
In fact, the friction law of Eq. (\ref{eq4'}) gives the tangential stress $\sigma_{nt}$ on the sphere, which can be integrated to yield the  force exerted by the medium on the cell in response to the spontaneous cortical flow. Assuming that the flow in the medium is much slower than the cortical flow, this force is non zero and reads $F=\frac{8\pi}{3}h_0R^2\xi v_0$, where $h_0$ is the cortex thickness and $v_0$ is the amplitude of the cortical flow,  showing that the instability of the mode $Y_{1,0}$ induces cell motion. Note that the flow produced in the  medium in response to cell motion induces a friction force $-F$, so that globally the cell acts on the fluid as a force dipole as expected.

Secondly, we note that this cortical instability is closely related to  bleb formation \cite{Salbreux2007,Charras2005,Charras2008,Tinevez:2009fk}. In the unstable regime, we find that the cortex is weakened, and could even lead to a hole, for instance at the  pole $\theta=\pi$ for the mode $Y_{1,0}(\theta)$. Above a certain activity threshold, the pressure inside the cell will be enough to break a weakened cortex, forming a bleb.  Larger modes $l>1$ correspond to several simultaneous blebs.  Our proposed mechanism of bleb formation is closely related to that of   \cite{Salbreux2007}, with the notable difference that their instabilities rely mainly on calcium dynamics.

Finally we compare these theoretical predictions to experiments conducted on MDA-MB-231 human breast adenocarcinoma cells seeded in  3D matrigel. These cells were found to invade and migrate in 3D matrigel with a striking round morphology, and to display neither membrane extension nor blebbing at the cell front, in contrast with classical experiments on flat substrates. 
Interestingly, silencing the main actin polymerization nucleator found in lamellipodial structures (Arp2/3) had no effect on the cell migration velocity, while it is known to be critical for migration on flat substrates. Moreover,  inhibition of myosin II using blebbistatin almost completely abolished cell movement indicating that myosin activity is essential for movement. This shows that an alternative motility mechanism, widely independent of actin treadmilling and dependent on myosin contractility, must be involved to trigger migration.  We suggest here that the cortical instability studied above, which relies mainly  on cortical contractility  could be the mechanism at work under these conditions.  High resolution confocal imaging of live MDA-MB-231 cells in matrigel  confirmed a highly non uniform distribution of cortical actin, which accumulates at the cell rear (see Fig.\ref{fig_geom} and \cite{Poincloux2010}) in agreement with the model. More precisely, live imaging analysis of actin cortical structures indicated the quantity  of cortical actin per perimeter length, quantified by $\rho$ in the model.  Figure (\ref{fig_donneintensite}) shows quantitatively the accumulation of actin at the back of the cell, and depletion at the front, in agreement with the predicted mode $Y_{1,0}$ which is unstable for $\zeta>\zeta_c$. Analysis of kymographs of these films provided the flow of cortical actin. Figure (\ref{fig_donnevitesse}) shows that the velocity profile obtained is also compatible with the mode $Y_{1,0}$,  which as we have shown induces cell motion. In addition, blocking beta1 integrin function with anti-human beta1 integrin monoclonal antibody 4B4 resulted in a strong reduction in migration and in the presence of blebs. In our model this corresponds to decreasing the friction parameter $\xi$ and consequently the threshold value $\zeta_c$, which leads to a shift of the 
instability towards the   higher modes $l>1$, corresponding to the formation of multiple simultaneous blebs. Together these experiments \cite{Poincloux2010} strongly suggest that in this example of cell migration in a 3D environment, motility is induced by the cortical instability described in the present paper.

To conclude, we have presented a generic model of cell motility generated by acto-myosin contraction of the cell cortex.  We identified analytically  dynamical instabilities  and showed that spontaneous cortical flow can induce cell migration in 3D environments.  Theoretical predictions compare well to experimental data of tumor cells migrating in 3D matrigel, and overall suggest that this contractility--based  mechanism of spontaneous cortical flows generation  could be a general mode of cell migration in 3D environments,  which strongly differs from the classical picture of lamellipodial motility.

%\bibliographystyle{vincent}
%\bibliography{activegels}

\end{document}